\documentclass[10pt, final, letterpaper, onecolumn, twoside]{IEEEtran}

\usepackage{cite}
\usepackage{graphicx}
\usepackage{subfigure}
\usepackage{amsmath}
\usepackage{amssymb}
\usepackage{epsfig}
\usepackage{times}

\IEEEoverridecommandlockouts

\begin{document}
\title{New Models for the Correlation in Sensor Data}
\author{\authorblockN{{\Large Samar Agnihotri}}\\
\authorblockA{Centre for Electronics Design and Technology\\
Indian Institute of Science, Bangalore-560012, India.\\
Email: samar@cedt.iisc.ernet.in}%
}
\maketitle

\begin{abstract}
In this paper, we propose two new models of spatial correlations in sensor data in a data-gathering sensor network. A particular property of these models is that if a sensor node knows in \textit{how many} bits it needs to transmit its data, then it also knows \textit{which} bits of its data it needs to transmit.
\end{abstract}

\section{Introduction}
\label{Intro}
In the past, \cite{Pradhan} and \cite{Cristescu} have proposed explicit models of spatial correlations in the sensor data. However, these models are not quite suitable in the context of the spatial correlations in the sensor data in the real data-gathering sensor networks, as the model proposed in \cite{Pradhan} is impractical and the model proposed in \cite{Cristescu} is computationally very intensive. Also, this latter model only gives the average number of bits transmitted by a node under a sensor polling schedule, more precisely, the differential entropy of a sensor node conditioned on the data of the nodes which have already transmitted their respective data. So, this model is not suitable if one wants to compute the number of bits transmitted by a node in the worst-case. Based on these limitations of the aforementioned models and the lack of any other non-trivial, practical model of spatial correlation in sensor data, in this work we propose two such models which are computationally simple, yet capture well our intuition of the spatial correlations in the sensor data. Also, these models can easily be used to compute both, the average and worst case number of bits transmitted by a node under a transmission schedule.

\section{New Models of Spatial Correlation}
Let $X_i$ be the random variable representing the sampled sensor reading at node $i\in \{1, \ldots, N\}$ and $B(X_i)$ denote the number of bits that the node $i$ has to transmit. Let us assume that each node $i$ has at most $n$ number of bits to transmit, so $B(X_i) = n$. However, due to the spatial correlation among sensor readings, each sensor may send less than $n$ number of bits. Let $d_{ij}$ denote the distance between nodes $i$ and $j$.

\begin{figure}[h]
\begin{center}
\includegraphics[angle=-90, width=8.85cm]{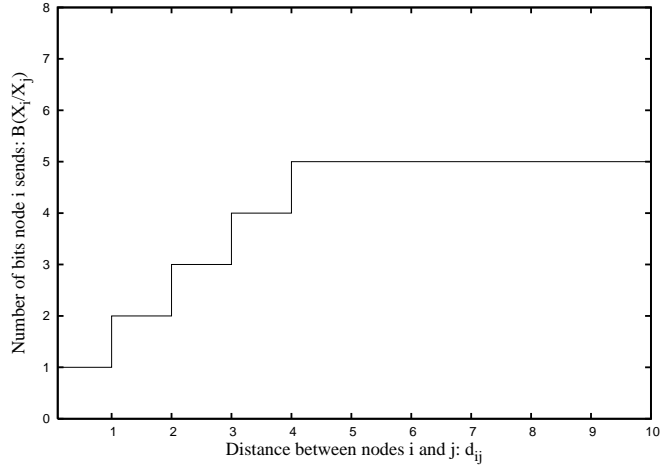}
\end{center}
\caption{First Data Correlation Model for $n=5$: plot of $B(X_i/X_j)$ defined in eqn \eqref{eqn:cor1} versus $d_{ij}$}
\label{fig1}
\end{figure}

\textbf{\textit{Model 1:}} Let us define $B(X_i/X_j)$, the number of bits that the node $i$ has to transmit when the node $j$ has already communicated its data, as follows:
\begin{equation}
\label{eqn:cor1}
B(X_i/X_j) = \left\{
                    \begin{array}{ll}
                     \alpha_1\lceil d_{ij}^{\beta_1} \rceil \mbox{ if } \alpha_1\lceil d_{ij}^{\beta_1} \rceil \le n,\\
                     n \mbox{ otherwise},
                    \end{array}
             \right.
\end{equation}
where the parameters $\alpha_1, \beta_1 \in \mathbb{R}, \alpha_1 > 0$, take care of the various application specific correlation effects. Figure \ref{fig1} illustrates this for $n=5, \alpha_1 = 1.0, \beta_1 = 1.0$. Here, it should be noted that when the node $j$ has already transmitted its data, then the node $i$ transmits no more than $B(X_i/X_j)$ bits of its $n$ bit data and we define these $B(X_i/X_j)$ bits to be the least significant $B(X_i/X_j)$ bits of its $n$ bit data. So, if the node $i$ knows in \textit{how many} bits it needs to transmit its data, then it also knows \textit{which} bits of its data it needs to transmit. Here we do not concern ourselves with how a node comes to know of in how many bits it should transmit its data. This discussion is beyond the scope of the present work and is discussed elsewhere \cite{submitted01}.

From the definition above in \eqref{eqn:cor1}, follows the symmetry of the conditional number of bits:
\begin{equation}
\label{eqn:cor_sym1}
B(X_i/X_j) = B(X_j/X_i)
\end{equation}

However, the definition of the correlation model is not complete yet and we must give the expression for the number of bits transmitted by a node conditioned on more than one node already having transmitted their bits. There are several ways in which this quantity can be defined. Here we have chosen to define it in the following two ways:
\begin{eqnarray}
B(X_i/X_1, \ldots, X_{i-1}) = \min_{1 \le j < i} B(X_i/X_j) \label{eqn:cor_cond_min} \\
B(X_i/X_1, \ldots, X_{i-1}) = \max_{1 \le j < i} B(X_i/X_j) \label{eqn:cor_cond_max}
\end{eqnarray}

So, according to equation \eqref{eqn:cor_cond_min}, the number of bits transmitted by node $i$ depends only on its nearest neighbor among all the nodes which have already communicated their data and according to \eqref{eqn:cor_cond_max}, it depends only on the farthest neighbor among all the nodes which have already communicated their data.

Let $\cal S$ be the set of nodes which have already transmitted their data. The rational behind the definition in \eqref{eqn:cor_cond_min} is that the sampled reading of a node is most correlated with the reading of its nearest neighbor in set $\cal S$. So, if two nodes are spatially close, then their data is most likely to differ only in the least significant bits. Similarly, the intuition behind the definition in \eqref{eqn:cor_cond_max} is that the sampled reading of a node is least correlated with the reading of its farthest neighbor in the set $\cal S$. So, the number of bits that a node has to transmit conditioned only on its farthest neighbor in set $\cal S$, gives the upper bound on the number of bits that the particular node has to transmit for the given set $\cal S$.

Note that when the nodes transmit their data according to some polling schedule $\pi$, then \eqref{eqn:cor1} denotes the number of bits transmitted by the node $\pi(i)$ when the node $\pi(j)$ has already transmitted its data. $B(X_i/X_1, \ldots, X_{i-1})$ in \eqref{eqn:cor_cond_min} and \eqref{eqn:cor_cond_max} should be interpreted similarly. Also note that for the correlation models in \eqref{eqn:cor_cond_min} and \eqref{eqn:cor_cond_max}, the sum of the number of bits transmitted by all the nodes depends on the transmission schedule according to which the nodes transmit their data.

\begin{figure}[h]
\begin{center}
\includegraphics[angle=-90, width=8.85cm]{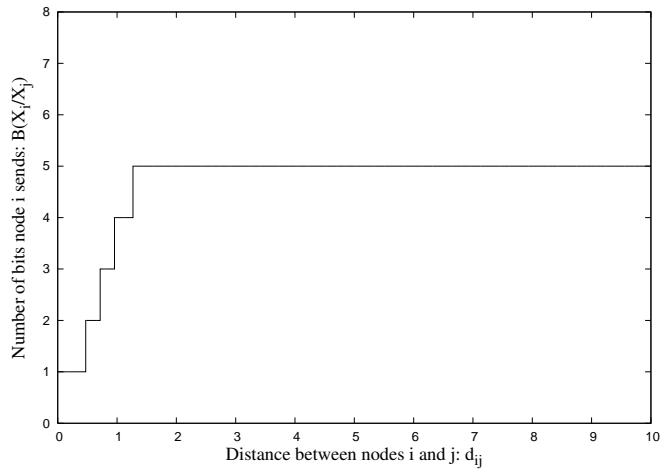}
\end{center}
\caption{Second Data Correlation Model for $n=5$: plot of $B(X_i/X_j)$ defined in eqn \eqref{eqn:cor2} versus $d_{ij}$}
\label{fig2}
\end{figure}

\textbf{\textit{Model 2:}} Let us generalize the previous model of spatial correlation in sensor data and define $B(X_i/X_j)$, the number of bits that the node $i$ has to transmit when the node $j$ has already communicated its data, as follows:
\begin{equation}
\label{eqn:cor2}
B(X_i/X_j) = \lceil n(1 - \alpha_2 e^{-\beta_2 d_{ij}^{2}}) \rceil,
\end{equation}
where the parameters $\alpha_2, \beta_2 \in \mathbb{R}, \alpha_2 > 0$, take care of the various application specific correlation effects. Figure \ref{fig2} illustrates this for $n=5, \alpha_2 = 1.0, \beta_2 = 1.0$. Once more, it should be noted that when the node $j$ has already transmitted its data, then the node $i$ transmits no more than $B(X_i/X_j)$ bits of its $n$ bit data and we define these $B(X_i/X_j)$ bits to be the least significant $B(X_i/X_j)$ bits of its $n$ bit data. So, if the node $i$ knows in \textit{how many} bits it needs to transmit its data, then it also knows \textit{which} bits of its data it needs to transmit.

For the small values (in magnitude) of the exponent on the right-hand side and $\alpha_2 = 1.0$, the correlation model in \eqref{eqn:cor2} reduces to the correlation model in \eqref{eqn:cor1}, if we identify $\alpha_1 = \beta_2$ and $\beta_1 = 2$. So, the model in \eqref{eqn:cor1} is the linear approximation of the model in \eqref{eqn:cor2}.

From the definition above in \eqref{eqn:cor2}, follows the symmetry of the conditional number of bits:
\begin{equation}
\label{eqn:cor_sym2}
B(X_i/X_j) = B(X_j/X_i)
\end{equation}

Now, let us define the number of bits transmitted by a node conditioned on more than one node already having transmitted their bits as:
\begin{equation}
\label{eqn:cor2_cond}
B(X_i/X_1, \ldots, X_{i-1}) = \Big\lceil n\Big(1 - \alpha \sum_{j = 1}^{i-1}e^{-\beta d_{ij}^{2}}\Big) \Big\rceil.
\end{equation}

$B(X_i/X_1, \ldots, X_{i-1})$ denotes the maximum number of bits that the node $i$ transmits, given that the nodes from set ${\cal S} = \{1, \ldots, i-1\}$ have already transmitted their data.

The intuition behind the above model of correlation is that the number of bits that a node has to transmit, with the nodes in the set $\cal S$ having already transmitted their data, should be the (weighted) average of number of the bits that the particular node has to transmit conditioned on all the node in the set $\cal S$ individually. The choice of the exponential dependence on the internode distance is based on the \textit{Gaussian} correlation model proposed in \cite{Cristescu}.

Note once more that when the nodes transmit their data according to some polling schedule $\pi$, then \eqref{eqn:cor2} denotes the number of bits transmitted by the node $\pi(i)$ when the node $\pi(j)$ has already transmitted its data. $B(X_i/X_1, \ldots, X_{i-1})$ in \eqref{eqn:cor2_cond} should be interpreted similarly. Also note that for the correlation model in \eqref{eqn:cor2_cond}, the sum of the number of bits transmitted by all the nodes depends on the transmission schedule according to which the nodes transmit their data.

\end{document}